# Ego-perspective enhanced fitness training experience of AR Try to Move game

**Chongyu Zhang**

Technical University of Munich, Munich, Germany chongyu.zhang@tum.de

chongyu.zhang@tum.de

**Abstract.** AR, a recent emerging technology, has been widely used in entertainment to provide users with immersive, interactive, and, sometimes, engaging experiences. The process of rehabilitation treatment and motor training process is often boring, and it is well known that users' exercise efficiency is often not as efficient as in a rehabilitation institution. Thus far, there is no effective upper limb sports rehabilitation training game based on the ego-perspective. Hence, with the objective of enhancing the enjoyment experience in rehabilitation and more effective remote rehabilitation training, this work aims to provide an AR Try to Move game and a convolutional neural network (CNN) for identifying and classifying user gestures from a self-collected AR multiple interactive gestures dataset. Utilizing an AR game scoring system, users are incentivized to enhance their upper limb muscle system through remote training with greater effectiveness and convenience.

## 1. Introduction

AR, a technology that superimposes virtual elements onto the physical world, has garnered substantial attention in recent times due to its characteristics, real and virtual combined, interactive and 3D spatial context [1]. As the demand for extended and monotonous rehabilitation training for post-stroke patients has grown [2], and with the rise of digital medicine to facilitate home-based training, the integration of AR technology provides a promising solution. Khademi et al. designed an AR program for post-stroke patients with upper-limb impairments. This program detects essential hand movements and offers tactile feedback interactively, allowing users to engage in activities pouring water, and more for a lifelike experience [3]. Chen et al. designed an AR game that integrates hand-controlled virtual football interaction with traditional Bobath therapy. The game incorporates virtual bone joint overlays and football movements onto the patient's video feed, facilitating the assessment of lower-limb balance [4]. In other research [5][6][7], various AR games were introduced with feedback systems. These games enabled users to engage in activities like car driving, board games, and Pokémon Go, with their performance tracked and showcased for remote transmission to the therapist. Existing AR applications exhibit the potential to harness the benefits of AR, enhancing interactivity and providing valuable guidance in monotonous training processes. However, prevalent AR games tend to suffer from limited design scope and overly simplistic gameplay. These limitations hinder their ability to offer users a truly immersive and ego experience, such as combining their musculoskeletal system information in real-time during rehabilitation training.

We want to enhance users' enjoyment and interactive engagement by leveraging multiple sensory channels (visual, auditory, and fitness-related stimuli), all while maintaining the same level of effectiveness as that achieved in a traditional rehabilitation institute. We selected Microsoft® HoloLens 2 HMD as an implementation unit, because of its good performance, and ease of development. To reach the goal above, this paper made the following contribution, and one pipeline is represented in Figure 1:

- design one AR puzzle game named Try to Move incorporated a variety of rehabilitation training interactive gestures, making it engaging and diverse for users to generate a wide range of upper limb movements.
- collect one user multiple interactive gestures AR game dataset, classify all multiple interactive gestures in 16 classes with a lighting CNN model achieved high accuracy, forming a scoring and reward system integrated with multiple factors in AR game.

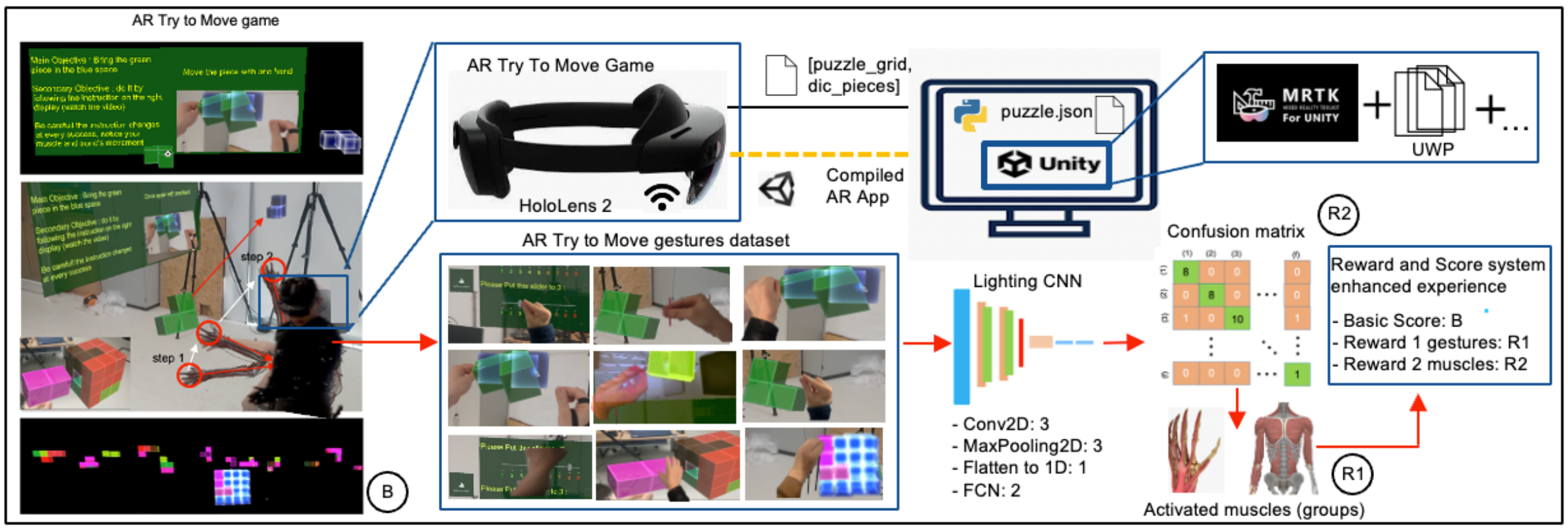

**Figure 1.** Pipeline of ego-perspective AR Try to Move game[1].

## 2. Methodology

As discussed earlier, in order to enhance the enjoyment of rehabilitation and make remote training more effective, this paper proposes in this section how to design the AR Try to Move game with more upper-limb rehabilitation conductive multiple multiple interactive gestures in visible areas and spatial movement in HoloLens 2, and the creation of lighting CNN for multiple interactive gestures classification. We discuss our development in 2 parts as follows.

### *2.1. Design for AR Try to Move game*

Based on Table 1 proposed rehabilitation process conducive multiple interactive gestures [8][9], this work aims to let users take those gestures more and reward users more with more interactive gestures and muscle activation. This work summarizes the mainly involved human upper-limb muscle (group) while performing the above movement and gestures based on previous research and extensive literature review. This work designed an AR Try to Move game (Figure 1) with four different levels of difficulty. Users need to complete our game by taking colorful pieces with different shapes into one blue target container. Every interaction by "taking" in our game will perform the multiple interactive gestures below. The AR game design part refers to this pipeline[2].

**Table 1. Propose rehabilitation conducive gestures and participating muscle (group)**

| Participating motor part | Main participating muscle (group) | N | Multiple interactive gesture |
|---|---|---|---|
| Upper/lower limb | Involved upper/lower limb body muscles | - | Move forward (1) and backward (2), turn right (3) and left (4) |
| Forearm, Humerus | Biceps Brachii, Triceps Brachii, Deltoid, Trapezius, Subscapularis, Subclavius, Teres Minor, Infraspinatus, Brachioradialis [10][11] | 9 | Upper and front arm folding movement (5) |

[1] This project package is available in GitHub: https://github.com/marschongyuzhang/mla-ar-try-to-move-game-
[2] Pipeline for puzzle game: https://github.com/Edouard99/MixedRealityPuzzleGame

| | | | |
|---|---|---|---|
| Forearm, Hand | Flexor Carpi Radialis, Flexor Carpi Ulnaris, Palmaris Longus, Flexor Digitorum Superficialis, Flexor Digitorum Profundus, Extensor Carpi Radialis Brevis, Extensor Carpi Radialis Longus, Extensor Carpi Ulnaris, Extensor Digitorum, Extensor Digiti Minimi [13][14][15] | 10 | Movement of the forearm drives movement of the wrist (6) |
| Forearm, Hand | Extensor Digitorum, Extensor Indicis, Extensor Digiti Minimi, Extensor Pollicis Longus, Extensor Pollicis Brevis, Extensor Carpi Radialis Longus [12][13][14][15] | 6 | Wrist extension (7) |
| Forearm, Hand | Flexor Digitorum Superficialis, Flexor Digitorum Profundus, Flexor Pollicis Longus, Flexor Carpi Radialis, Flexor Carpi Ulnaris, Palmaris Longus [12][13][14] [15] | 6 | Wrist flexion (8) |
| Forearm, Hand | Extensor Digitorum, Extensor Indicis, Extensor Digiti Minimi, Extensor Pollicis Longus, Extensor Pollicis Brevis, Extensor Carpi Radialis Longus, Extensor Carpi Ulnaris, Extensor Digitorum Communis [12][16][17] | 8 | Open hand (9) |
| Forearm, Hand | Flexor Digitorum Superficialis, Flexor Digitorum Profundus, Flexor Pollicis Longus, Flexor Carpi Radialis, Flexor Carpi Ulnaris, Palmaris Longus, Flexor Pollicis Brevis, Flexor Digiti Minimi Brevis [12][16][17] | 8 | Close hand (10) |
| Forearm, Hand, Finger | Flexor Digitorum Superficialis, Flexor Digitorum Profundus, Extensor Digitorum, Extensor Indicis, Interossei Muscles, Lumbrical Muscles, Thenar and Hypothenar Muscles, Flexor Carpi Radialis, Extensor Carpi Radialis Longus [12][16][17][18] | 10 | Tap with index-finger (a) |
| Forearm, Hand, Finger | Flexor Digitorum Superficialis, Flexor Digitorum Profundus, Flexor Pollicis Longus, Flexor Carpi Radialis, Flexor Carpi Ulnaris, Thenar Muscles, Hypothenar Muscles, Lumbrical Muscles, Interossei Muscles, Extensor Muscles [12] [16][17][18] | 10 | All-finger grasping (b) |
| Forearm, Hand, Finger | Thenar Muscles, Lumbrical Muscles, Interossei Muscles, Flexor Pollicis Longus, Flexor Digitorum Superficialis, Flexor Digitorum Profundus, Extensor Muscles, Opponent Muscles [12][16][17] | 8/ 16 | Index-Thumb-finger grasping (single hand (c) and double hands (d)) |
| Humerus, Forearm | Biceps Brachii, Supinator [13][16] | 2 | Turn the palm upwards (e) |
| Humerus, Forearm | Pronator Teres, Pronator Quadratus, Brachioradialis [13][16] | 3 | Turn the palm downwards (f) |

### *2.2.1 Generate random puzzle*

First, we randomly generate the number of pieces, the shape, and the size of the target container required for game design according to the following algorithm 1 in Figure 2 in Python. One colorful piece is a combination of 3D squares. Every piece's name and position are jointly defined in a matrix (e.g. piece 2 with 2 squares defines all 3D matrix locations where the number 2 can be found). To imitate the randomness of the piece appearance in the real 3D puzzle, realize the pre-defined multiple interactive gestures and spatial movement, means that: 1.) the location and orientation of pieces are random adjacent to each other along the *Unity* axis, 2) all pieces should be separately located in different positions in the AR environment. The random puzzle grid with pieces dictionary for different levels of 4 difficulty is generated. It takes the input puzzle size S and number of pieces N for locations in a 3D matrix filled with N (simultaneously as "name of the pieces"). As iteration begins, the algorithm 1 tries to extend to a free adjacent location in the next step. For any free location with the true label, the extension is a success, and the number of the piece N replaces the last value. After that, if still not all locations are successfully placed (in case of *success_2_cube == True*), it calculates the number of missing pieces that need to be added to the puzzle grid to ensure complexity. The JSON files with output *puzzle_grid* and *dic_pieces* are generated for game design in *Unity*.

**Algorithm 1** Algorithm for puzzle generation puzzle.json

```
Require: Puzzle size S, Number of pieces N
function GENERATE PUZZLE(S, N)
Initialization: dic_pieces
success_2_cube ← False
while i = 0 to 2 do
    Initialization: mat_preselect, index_start_pieces
    for k in N, index in index_start_pieces do
        mat_preselect[index] ← k
    end for
    success_2_cube ← True
    puzzle_grid ← size_Puzzle
    for k in N, index in index_start_pieces do
        position = index/(S^2)
        Find neighbor puzzle_grid
        dic_pieces ← [position, target_position]
    else
        success_2_cube ← False
    end for
    Case: success_2_cube == True
    Do: puzzle_grid, dic_pieces ← calculate missing
end while
Output: puzzle_grid, dic_pieces
```

```
function FIND NEIGHBOR(matrix, index)
for i = 0 to 2 do
    index_forward ← index + (i, 1, 0)
    index_backward ← index − (i, 1, 0)
    if index_forward within boundary then
        neighbor ← 1 "available"
    end if
    if index_backward within boundary then
        neighbor ← 1 "available"
    end if
end for
```

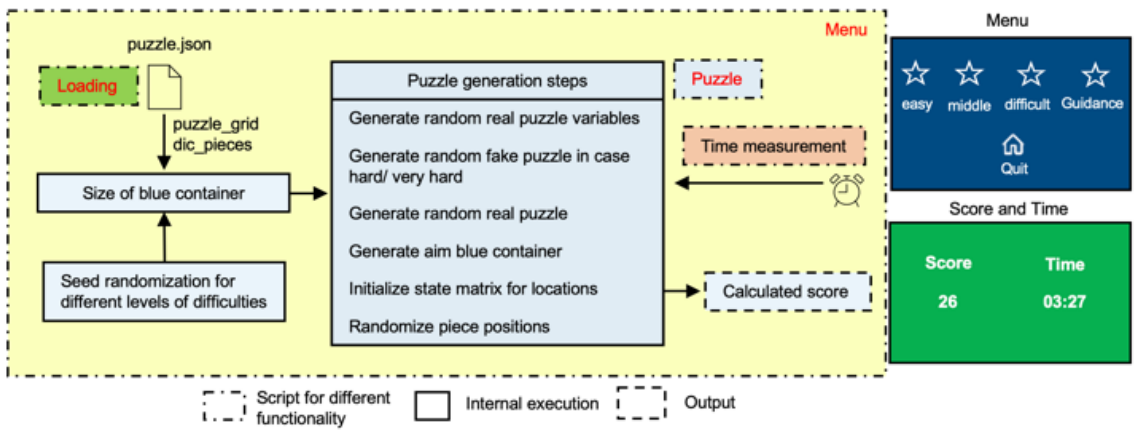

**Figure 2.** Pipeline of ego-perspective AR Try to Move game, Algorithm 1 contains the function GENERATE PUZZLE and FIND NEIGHBOR, the right below part shows the AR game structure.

### *2.2.2 Design Try to Move game in Unity*

To enable users to move themselves and achieve multiple upper-limb interactions in a more realistic and fantastic real-world environment, we designed an AR Try to Move game in *Unity* for compilation in HoloLens 2. In *Unity*, the main structure, execution process, menu, and scoring board are implemented and shown in Figure 2. Once one button of the menu is activated, it then reads the generated *puzzle.json* file containing the dimension *puzzle_grid* and vector in puzzle dictionary *dic_pieces* first. With randomized seeds equivalent for presenting the different difficulties, a random number of colorful pieces and a blue puzzle container are randomized and generated through the puzzle generation steps in the puzzle function (grey block in Figure 2). While playing, the score is calculated with the Formula in Table 2 after finishing the puzzle game or time running out by time measurement.

When users want to increase the difficulty of our AR Try to Move game, they can freely increase the volume of the container, whose current size is 4x4x4, or change the seed number of generated puzzle pieces with their needs, making the game more personalized and diverse in style. The common issues in game design with faster gameplay and higher score rewards time by time due to experience can be solved with our design approach.

### *2.2. Classification of the multiple interactive gestures with CNN*

Based on our scoring system design, users will get a basic score about how well they have completed the game with limited time when the game is over. In addition, we obtained 2 reward scores based on the times of multiple interactive gestures and the number of muscle activations during the game, which is estimated by a lighting CNN. For rewards estimation, we designed one lighting CNN and a confusion matrix. CNN performs well in image processing. Therefore, a lighting CNN can complete the task of identifying and classifying multiple interactive gestures with high accuracy for our collected dataset with 16 classes of multiple interactive gestures in our game. Our CNN consists of three convolutional layers for upper limb feature extraction, followed by max-pooling layers for downsampling, a flattening layer, and two fully connected layers for classification, which have been presented in Figure 1.

## 3. Experiment

To verify that our game can allow users to take more upper limb multiple interactive gestures for more effective and enjoyable training, we recorded the complete situation of users when using HoloLens 2 to experience different levels and made it into our AR interactive gesture dataset. It contains all the gestures of 16 times game-playing, which first three times of 4 levels record the situation that users can obtain all 3 scores. The last time recording of 4 levels only contained the user's basic score.

Using our dataset we trained our CNN with a learning rate of $10^{-3}$, Adam as Optimizer and sparse categorical cross-entropy loss. We recorded after training the confusion matrix, which presents the times of the users' multiple interactive gestures ((1)-(10) and (a)-(f)) mentioned in Table 2 during playing as R1 in its diagonal, the number of participating muscle activations involved in each interactive gesture (R2). The classification accuracy after training reached 98.7% or higher, which resulted in a total of no more than 5 misclassified multiple interactive gestures in the nondiagonal position in every confusion matrix. The calculation of the basic score follows the guideline: within the specified time, the less time it takes, the higher the basic score will be. Due to our game design with random generalization, higher scores due to higher proficiency are effectively avoided. Summarized with the confusion matrix of two rewards, a final score (F) came out.

**Table 2. Comparative results: recording of B and classification for R1, R2, and calculation F for 16 classes of upper limb multiple interactive gesture**

| Level | Time | B | Times of multiple interactive gesture* (Hand + Arm) w.r.t Table 1 | | | | | | | | | | | | | | | | R1 | R2 | F[1] |
|---|---|---|---|---|---|---|---|---|---|---|---|---|---|---|---|---|---|---|---|---|
| | | | (1) | (2) | (3) | (4) | (5) | (6) | (7) | (8) | (9) | (10) | (a) | (b) | (c) | (d) | (e) | (f) | | | |
| Guidance | 186s | - | 1 | 1 | 6 | 6 | 6 | 6 | 6 | 6 | 0 | 0 | 4 | 0 | 6 | 0 | 0 | 0 | 48 | 274 | 322 |
| Guidance | 237s | - | 1 | 1 | 6 | 6 | 6 | 6 | 8 | 7 | 0 | 0 | 4 | 0 | 4 | 2 | 0 | 1 | 52 | 311 | 363 |
| Guidance | 221s | - | 1 | 0 | 5 | 6 | 6 | 6 | 7 | 6 | 0 | 0 | 4 | 0 | 4 | 2 | 0 | 0 | 47 | 296 | 343 |
| Guidance | 230s | - | - | - | - | - | - | - | - | - | - | - | - | - | - | - | - | - | 32 | 254 | 286 |
| Easy | 93s | 61 | 8 | 8 | 9 | 14 | 10 | 13 | 0 | 3 | 2 | 0 | 6 | 0 | 6 | 4 | 0 | 0 | 83 | 426 | 570 |
| Easy | 102s | 57 | 6 | 6 | 7 | 7 | 8 | 9 | 2 | 6 | 1 | 0 | 6 | 0 | 6 | 5 | 1 | 2 | 72 | 414 | 543 |
| Easy | 87s | 64 | 6 | 6 | 9 | 9 | 11 | 12 | 0 | 6 | 1 | 1 | 6 | 0 | 6 | 4 | 0 | 0 | 77 | 443 | 584 |
| Easy | 105s | 56 | - | - | - | - | - | - | - | - | - | - | - | - | - | - | - | - | 59 | 378 | 493 |
| Middle | 147s | 69 | 7 | 7 | 10 | 10 | 9 | 12 | 3 | 8 | 5 | 2 | 9 | 0 | 8 | 8 | 1 | 0 | 99 | 607 | 775 |
| Middle | 162s | 66 | 11 | 11 | 13 | 15 | 13 | 16 | 5 | 7 | 6 | 0 | 8 | 0 | 11 | 6 | 0 | 0 | 122 | 661 | 849 |
| Middle | 158s | 67 | 8 | 8 | 10 | 9 | 11 | 14 | 1 | 9 | 5 | 3 | 8 | 0 | 9 | 5 | 0 | 1 | 101 | 598 | 766 |
| Middle | 167s | 65 | - | - | - | - | - | - | - | - | - | - | - | - | - | - | - | - | 87 | 522 | 674 |
| Difficult | 246s | 59 | 10 | 10 | 15 | 15 | 16 | 22 | 2 | 13 | 10 | 3 | 13 | 2 | 15 | 7 | 2 | 0 | 155 | 944 | 1158 |
| Difficult | 218s | 64 | 15 | 15 | 17 | 17 | 15 | 23 | 4 | 12 | 12 | 2 | 12 | 3 | 17 | 6 | 1 | 2 | 173 | 963 | 1200 |
| Difficult | 255s | 58 | 13 | 13 | 18 | 18 | 19 | 20 | 3 | 15 | 12 | 2 | 13 | 2 | 14 | 7 | 3 | 3 | 175 | 980 | 1213 |
| Difficult | 260s | 57 | - | - | - | - | - | - | - | - | - | - | - | - | - | - | - | - | 147 | 902 | 1106 |

[1]: Formula (1) for F calculation: $F = \frac{100}{t_{total}} * (t_{total} - t_{end}) + \sum_{t=0}^{t_{end}} Times\ of\ gesture + \sum_{t=0}^{t_{end}} Times\ of\ gesture * N_{number\ of\ participating\ muscle\ activations}$

[*]: Times of multiple interactive gesture (Hand + Arm) are classified with CNN and here is the classification results of the diagonal of confusion matrix.

Every time of playing the game can collect around 50, 90, 120, and 180 pictures of the user's upper limb interactive gestures and spatial movements. In guidance, users need to follow the digital instructions to complete the game and obtain rewards. As the difficulty level increased, the clearance time became longer with the appearance of fake pieces, and the game playing became more enjoyable. Users always try more multiple interactive gestures to complete a higher level of difficulty and deepen their understanding of their muscle activation states. We also found that the time-based score calculation principle can encourage users to complete the game by more interacting in the shortest possible time because more multiple interactive gestures and less time meant a higher final score. The comparative groups in Table 2 were not classified by CNN. We only quietly recorded the number of users' multiple interactive gestures and muscle activations and calculated the F, without providing any information to users. We discover that through obtaining two rewards, users can be motivated to interact more, activate their muscles, take more gestures, and with higher enjoyable experience. Therefore, our research can be used for digital rehabilitation training at home. The same rehabilitation training effect can be expected because of our game and reward system. All analysis was meant in line with our original design intentions for the AR Try to Move game.

## 4. Conclusion and future work

In summary, we proposed in our work one AR Try to Move game with CNN classification and reward system to let users achieve the most interactive gestures in the game and the training process. Our games are designed based on 16 multiple interactive gestures that are conducive to rehabilitation training. The lighting CNN can classify the recorded multiple interactive gestures taken by the users with high accuracy and provide a confusion matrix every time. Based on our design, calculating, and informing users of basic scores and rewards allows users to take more interactive gestures and train themselves more. The experiment shows that our game mechanism also encourages the user to take more multiple interactive gestures and spatial movements to complete the game in as little time as possible and achieve a final score as high as possible, carry out more enjoyable and interesting upper limb training with a better understanding of their own muscle activations in movement with HoloLens 2. More effective upper limb rehabilitation training can be expected to take remotely even in non-medical centers.

We discovered one possibility for future research. Post-stroke patients often cannot effectively move themselves independently. Therefore, our expectation is to ensure the configuration of stimulation signals tailored to various upper limb gestures. It empowers users to pinpoint the specific muscle activation and stimulation location, thereby facilitating our game completion with stimulation assistance. This versatility broadens the scope of our software, aligning it with the diverse rehabilitation requirements of patients with various medical conditions for digital medicine.